\documentclass{jetpl}
\twocolumn \lat

\title{On an explicit construction of Parisi landscapes in finite dimensional Euclidean spaces}
\rtitle{Parisi landscapes in Euclidean spaces}

\author{Yan V Fyodorov$^{1,2}$ and Jean-Philippe Bouchaud$^{3,4}$}

\address{ $^1$ Institut f\"{u}r Theoretische Physik,
Universit\"{a}t zu K\"{o}ln, 50937 K\"{o}ln, Germany \\$^2$ School
of Mathematical Sciences, University of Nottingham, Nottingham
NG72RD, England\\$^3$ Science \& Finance, Capital Fund Management
6-8 Bd Haussmann, 75009 Paris, France
\\$^4$ Service de Physique de l'{\'E}tat Condens{\'e}
Orme des Merisiers -- CEA Saclay, 91191 Gif sur Yvette Cedex,
France}

\abstract{We construct a $N-$dimensional Gaussian landscape with
multiscale, translation invariant, logarithmic correlations and
investigate the statistical mechanics of a single particle in this
environment.  In the limit of high dimension $N\to \infty$ the
free energy of the system in the thermodynamic limit coincides
with the most general version of Derrida's Generalized Random
Energy Model. The low-temperature behaviour depends essentially on
the spectrum of length scales involved in the construction of the
landscape. We argue that our construction is in fact valid in any
finite spatial dimensions $N\ge 1$.}


\begin{document}
\maketitle

\newcommand \be  {\begin{equation}}
\newcommand \bea {\begin{eqnarray} \nonumber }
\newcommand \ee  {\end{equation}}
\newcommand \eea {\end{eqnarray}}

The idea of energy landscapes pervades the theoretical description
of glasses, disordered systems, proteins, etc. \cite{Wales}. The
general goal is to classify typical random potentials and
establish their universal properties, not unlike the Random Matrix
Theory paradigm. This knowledge can then hopefully be used to
describe generic static and dynamic properties of complex systems, by
addressing single point particle behavior in such potentials. In
this respect, the Parisi solution for spin-glasses is fascinating:
it reveals landscapes with a surprisingly complex, hierarchical
structure of valleys within valleys within valleys, etc. \cite{Parisi}.
It is often however argued that the ultrametric
properties of Parisi landscapes are hardly compatible with a
finite dimensional, translation invariant space.

In this paper we provide an explicit construction of a Gaussian
random potential in Euclidean, $N$ dimensional spaces, with a
specific form of long-ranged correlations which reproduces all the
features of Parisi landscapes. More precisely, we show that the
thermodynamics of a single particle in a multiscale,
logarithmically correlated potential is exactly described by
Derrida's Generalized Random Energy Model (GREM, \cite{GREM}),
with an arbitrary (possibly infinite) number of levels of
hierarchy. Although our proof concerns, strictly speaking, the
limit $N \to \infty$, we are confident that our results hold in
arbitrary finite dimension $N \geq 1$. This conviction is built
both on physical arguments and on the beautiful results of
Carpentier and Le Doussal \cite{CLD} on the monoscale version of
our model in finite dimensions, which, as shown recently, match
the exact results of the same model when $N \to \infty$ \cite{FS}.
The model is defined as follows: the position of the particle,
confined inside an $N-$dimensional spherical box of radius $L$, is
described by the coordinate vector ${\bf r}=(r_1,...,r_N),\,\,
|{\bf r}|\le L$. It feels a Gaussian-distributed random potential
$V({\bf r})$ with zero mean, and with covariance chosen to be
isotropic, translation invariant and with a well-defined large
$N-$limit:
\begin{equation}\label{potential}
\left\langle V\left({\bf r}_1\right) \, V\left({\bf
r}_2\right)\right\rangle_V=N\,f\left(\frac{1}{2N}({\bf r}_1-{\bf r}_2)^2\right)\,.
\end{equation}
In Eq.(\ref{potential}) and henceforth the notation
$\left\langle\ldots\right\rangle_V$ stands for an ensemble average
over the random potential, and $f$ is a well behaved function of
order unity. The thermodynamics of this model is described by the
free energy:
\begin{equation}\label{freeendef}
F_N=-\beta^{-1}\,\langle \ln{Z(\beta)}\rangle_V ,\, \quad
Z(\beta)=\int_{|{\bf r}|\le L} \exp{-\beta V({\bf r})}\, d {\bf
r}\,
\end{equation}
as a function of the inverse temperature $\beta={1}/{T}$.

Models of this kind has been studied extensively, and in the
high-dimensional limit detailed analytical calculations performed
in \cite{MP} revealed that the nature of the low temperature phase
is essentially dependent on the behavior of the covariance $f(u)$
at large distances. Namely, for short-ranged correlated
potentials, the low temperature phase turns out to be described by
one-step replica symmetry breaking scheme of Parisi. In contrast,
for the case of long-ranged correlated potentials with $f(u)$
growing as $u^{2\gamma}$, the full infinite-hierarchy replica
symmetry breaking (FRSB) scheme has to be used.

The problem was reconsidered in much detail recently in \cite{FS}.
In the limit $N \to \infty$, one actually finds a true phase
transition as a function of temperature provided the size of the
confining sphere $L$ is scaled as $R \sqrt{N}$, with the parameter
 $0<R<\infty$ playing the role of effective radius of the sample.
As the existence of a phase transition for finite-size systems is
a kind of pathology of the infinite-dimensional approximation, one
is mainly interested in the thermodynamic limit $R\to \infty$. A
simple analysis then reveals a special role played by {\it
logarithmic} correlation function:
\begin{equation}\label{2c}
f(u)=f_0-g^2\ln{(u+a^2)}\,,
\end{equation}
where $g$, $a$ and $f_0$ are given constants. In particular, only
for such case the critical temperature $T_c$ tends to a finite
value: $ T_c(R\to \infty)=g$. Furthermore, the free-energy found
in this limit is given by the well-known Random Energy Model
expression \cite{REM}. Interestingly, these results coincide {\it
precisely}, up a trivial rescaling, with those obtained earlier
for the same potential (\ref{2c}) at $1\le N<\infty$ dimensions
from a Renormalisation Group treatment \cite{CLD}. In the standard
interpretation, below $T_c$ the partition function becomes
dominated by a few sites with particularly low random potential,
where the particle ends up spending most of its time \cite{MB}.
Note that the logarithmic growth of the variance of the potential
appears naturally in various physical systems of diverse nature,
see \cite{CLD,2d}.

The main observation of the present paper is that the above
picture, despite looking rather complete, still misses a rich
class of possible behavior that survives in the thermodynamic
limit $R \to \infty$. Namely, given any increasing positive
function $\Phi(y)$ for $0<y<1$, we demonstrate below that if one
considers potential correlation functions $f(u)$ which take the
following scaling form
\begin{equation}\label{scalingln}
f(u)=-2 \ln{R}\,\, \Phi\left(\frac{\ln{(u+a^2)}}{2\ln{R}}\right),
\quad 0 \le u<R^2,
\end{equation}
the thermodynamics of our system in the limit $R\to \infty$ is
precisely equivalent to that of celebrated Derrida's Generalized
Random Energy Model (GREM)\cite{GREM}. The REM-like case
Eq.(\ref{2c}) turns out to
 be only a (rather marginal) representative of
this class: $\Phi(y)=g^2 y$.

Let us explain the motivation of the above form, which will make
the physical interpretation of the results quite transparent. The
idea is to write $V({\bf r})$ as a (possibly infinite) sum of
independent Gaussian potentials: $ V({\bf r})=\sum_{i=1}^K
V_i({\bf r}) $, each with a covariance: \be \left\langle V_i
\left({\bf r}_1\right) \, V_k \left({\bf
r}_2\right)\right\rangle_V=\delta_{i,k} N\,f_i \left(\frac{1}{2N}
({\bf r}_1-{\bf r}_2)^2\right)\,, \ee where
$f_i(u)=-g_i^2\ln{(u+a^2+a_i^2)}\,$ as in (\ref{2c}), but each
with its own constant $g_i$, and small-scale cutoffs $a_i$ chosen
to grow as a power-law of the system size: $a_{i}=R^{\nu_i}$ with
$0 \leq \nu_i \leq 1$. Taking the continuum limit $K \to \infty$
with a certain density $\rho(\nu)$ of exponents $\nu_i$, we end up
with:
\begin{equation}\label{scalingln1}
f(u)=-\int_{0}^{1}\rho(\nu) g^2(\nu)
\ln{\left(u+a^2+R^{2\nu}\right)}\,d\nu, \quad 0\le u\le R^2.
\end{equation}
Now, introducing $u+a^2 \equiv R^{2y}$ and identifying with Eq.
(\ref{scalingln}) in the $R \to \infty$ limit, we obtain that the
function $\Phi$ has the following representation: \be
\label{JP1}\Phi(y) = y \int_0^y \rho(\nu) g^2(\nu) \,d\nu +
\int_y^1 \nu \rho(\nu) g^2(\nu) \,d\nu. \ee  Note also that in
this representation, $\Phi'(y)= \int_0^y \rho(\nu) g^2(\nu) \,d\nu
\geq 0$, and $\Phi''(y) \geq 0$, where the number of dashes here
and below indicates the number of derivatives taken. The main
result of this work is the following: depending on the nature of
the spectrum of the exponents $\nu$, discrete or continuous, we
will recover, in the thermodynamic limit, either the free energy
of the original GREM with discrete hierarchical structure, or of
its continuous hierarchy analogue (see (\ref{freeenfin}) below)
analysed recently in much detail by Bovier and Kurkova \cite{BoK},
see also \cite{PWW}.

The physical interpretation of our results is as follows. Instead
of one localisation transition temperature $T_c$ at which the
particle chooses a finite number of ``blobs'' of size $O(a)$ where
the potential is particularly deep, there appears $K$ different
transition temperatures, where the particle localizes on finer and
finer length-scales. The largest transition temperature $T_1$
corresponds a condensation of the Boltzmann weight inside a few
blobs of large size $O(R)$, but the particle is still completely
delocalized {\it inside} each blob. As the temperature is reduced,
the REM condensation takes place over smaller blobs of size
$O(R^\nu)$ inside each already occupied large blobs, and this
scenario repeats itself as the temperature is reduced, each time
``zooming" in on a smaller scale \cite{note}.

The equilibrium free energy per degree of freedom of our model,
$F_{\infty}=\lim_{N\to\infty}F_N/N$ with $F_N$ defined in Eq.
(\ref{freeendef}), can be found in a standard way for any
covariance $f(u)$ using the replica trick.  The details of the
corresponding analysis can be found in \cite{FS}, and we give
below a summary of the most essential formulae for the FRSB
situation. For finite $R$, the low temperature phase is
characterised by the existence of a nontrivial, non-decreasing
function $x(q),\, q\in [q_0,q_k]$, with the two parameters $q_0$
and $q_k$ satisfying the inequality $0\le q_0\le q_k\le q_d \equiv
R^2$. The corresponding $F_{\infty}$ can be written in terms of
only those two parameters, see Eq.(58) of \cite{FS}. Here we
choose instead to introduce, along the line of the physical
discussion given above, two characteristic ``blob" sizes (actually
size squared) $d_{\min}=R^2-q_k,\,d_{\max}=R^2-q_0$ in terms of
which:
\begin{eqnarray}\label{freeenglass}
&&F_{\infty}=\frac{1}{2T}\left[f(d_{\min})-f(0)-d_{\min}f'(d_{\min})\right]
\\ \nonumber&&-\frac{T}{2}\ln{\left[2\pi e
d_{\min}\right]}+\frac{f'(d_{\max})}{\sqrt{f''(d_{\max})}}-\int_{d_{\min}}^{d_{\max}}\sqrt{f''(u)}\,du,
\end{eqnarray}
where $d_{\min} \leq d_{\max}$ can be found
for a given temperature $T$ from the equations
\begin{equation}\label{minmax}
0 \leq d_{\min}=\frac{T}{\sqrt{f''(d_{\min})}},\quad
d_{\max}=R^2+\frac{f'(d_{\max})}{f''(d_{\max})} \leq R^2
\end{equation}
Finally, the Parisi order-parameter function, which takes the values between $0$ and $1$ and is the main measure of the
ultrametricity in the phase space, has the following shape
\begin{equation}\label{con4}
x(d)=-\frac{T}{2}\frac{f'''(d)}{[f''(d)]^{3/2}},\quad \forall
d\in[d_{\min},d_{\max}]\,\,.
\end{equation}
 where we performed the overall change $q\to d=R^2-q$ in comparison with
 \cite{FS}. This function must be now {\it non-increasing}, and one can verify that this is precisely
the case e.g. for the family $f(u)$ in Eq.(\ref{scalingln}).

The above solution is valid for the temperature range $0\le T\le
T_{c}$, where the critical temperature $T_{c}$ is given in terms
of the largest blob size $d_{\max}$ as:
\begin{equation}\label{AT}
T_{c}=d_{\max}\sqrt{f''(d_{\max})}\, .
\end{equation}
Above this temperature the solution is replica-symmetric (RS),
corresponding to a
 delocalized phase for the particle: no particular region dominates the
partition function. The corresponding free energy is given by:
\begin{eqnarray}\label{freeensym}
F_{\infty}=-\frac{T}{2}\ln{\left[2\pi
d_{s}\right]}+\frac{1}{2T}\left[f(d_s)-f(0)\right]-\frac{T}{2}\frac{R^2}{d_s}
\end{eqnarray}
where $d_s$ satisfies
\begin{equation}\label{repsym}
d_{s}=R^2+\frac{d^2_{s}}{T^2}f'(d_{s})\, .
\end{equation}

We now consider specifically correlation functions $f(u)$ of the
form (\ref{scalingln}). In what follows we will use the convenient
notations $z=({2\ln{R}})^{-1}$ and $y=z\ln{(u+a^2)}$. As noted
above, our multiscale logarithmic model ensures that  $\Phi'(y)\ge
0$ and $\Phi''(y)\ge 0$ for any $0<y<1$. We will assume for
simplicity $\Phi'(0)=0$, relegating consideration of the general
case to an extended publication \cite{prep}. We start our analysis
assuming the function $\Phi''(y)$ is finite and differentiable,
but later on will relax this condition. Our first goal is to find
the largest blob size $d_{\max}$ from second equation in
Eq.(\ref{minmax}), and then to determine the critical
 temperature $T_{c}$. Introducing the scaling variable
$y_{\max}=z\ln{(d_{\max}+a^2)}$, in the thermodynamic limit $z\to
0$ we can look for a solution $y_{\max}(z)$ as a power series of
$z$. One immediately checks that $y_{\max}(z)=1-z\ln{2}+O(z^2)$.
This implies that the largest blob size is of the order of the
system radius: $d_{\max} \approx {R^2}/{2}\gg a^2$ for $R\to
\infty$.  Eq.(\ref{AT}) then yields the critical temperature given
in the thermodynamic limit by a very simple expression $
T_{c}=\sqrt{\Phi'(1)}$. Physically, at $T_c$, the sample breaks up
into blobs of size $o(R)$ and only a finite number of these blobs
are visited by the particle. However, within each blob, all sites
are more or less equivalent. Now we can treat along the same lines
the first equation in (\ref{minmax}) to determine the smallest
blob size $d_{\min}$ for $T < T_c$. It can again be conveniently
written in terms of the scaling variable
$y_{\min}=z\ln{(d_{\min}+a^2)}$. In the thermodynamic limit $z\to
0$, it is again natural to look for a solution $y_{\min}$ as a
power series of $z$, in which we only retain the first two terms:
$y_{\min}=\nu_*+cz+O(z^2)$. Due to our assumption on
differentiability of the function $\Phi'(y)$ we expand around
$y=\nu_*$, and after a simple calculation find $c=1$. This means
that $d_{\min}$ behaves like $d_{\min}=eR^{2\nu_*}$ for $R\to
\infty$, where $\nu_*$ satisfies the equation
\begin{equation}\label{main1}
T^2=\Phi'(\nu_*)\,.
\end{equation}
Since the function $\Phi'(y)$ is monotonously increasing for
$y>0$, and $\Phi'(0)=0$ we find that in the limit $R\to \infty $
(i.e. $z\to 0$), the equation Eq.(\ref{main1}) must have a unique
solution $0<\nu^*(T)<1$ in the range of temperatures $0<T <
T_{c}=\sqrt{\Phi'(1)}$. In this regime, $d_{\min} \ll d_{\max}$.
Physically, sites within blobs of size $d_{\min}$ or smaller are
not resolved by the particle, which visits all of them more or
less equally.

Now we can easily find the free energy $F_{\infty}$ by
substituting these results to Eq.(\ref{freeenglass}) and
extracting the leading term in the thermodynamic limit $z\to 0$.
We find the equilibrium free energy to be of the form
${F_{\infty}}=(\ln R) \,{\cal F}$, where for $0\le T\le T_{c}$
\be\label{freeenfin} -{\cal F}=
T\nu_*(T)+\frac{\left[\Phi(\nu_*)-\Phi(0)\right]}{T}
+2\int_{\nu_*}^1\sqrt{\Phi'(y)}\, dy \,. \ee  For $T>T_{c}$ the
solution of (\ref{repsym}) in the limit $R\to \infty$ is given by
$d_s=R^2\frac{T^2}{T^2+T_{c}^2}$ and substituting this to
(\ref{freeensym}) we find that the free energy is given by:
\begin{eqnarray}\label{freeensyma}
&& -{\cal F}=T+\frac{\left[\Phi(1)-\Phi(0)\right]}{T}\,.
\end{eqnarray}

Last but not least, we can determine the thermodynamic limit of
the order-parameter function $x(d)$ given by Eq.(\ref{con4}),
which determines in a precise way how the particle localizes on
different scales. To leading order in $z$ we find
$f''(u)=\Phi'(y)/(u+a^2)^2,\,f'''(u)=-2\Phi'(y)/(u+a^2)^3$ with
$y=z\ln{(u+a^2)}$. Introducing again the scaling variable
$\nu=\frac{\ln{(d+a^2)}}{2\ln{R}}$ for $d\in[eR^{2\nu_*},R^2/2]$
we see that the order-parameter function assumes the limiting
form:
\begin{equation}\label{con4a}
x(\nu)=\frac{T}{\left[\Phi'\left(\nu\right)\right]^{1/2}},\quad \forall \nu\in[\nu_*,1]\,\,.
\end{equation}

This completes our solution of the problem for the case of
continuous function $\Phi'(y)$. At this point it is rather
informative to consider the case of a discrete spectrum of $K$
exponents $\nu_{i},\,i=1,\ldots,K$ satisfying
$0<\nu_K<\nu_{K-1}<\ldots<\nu_{1}<\nu_0=1$. This corresponds to
$K$ superimposed logarithmic potentials with
\begin{equation}\label{disc}
g^2(\nu) \rho(\nu)=\sum_{i=1}^K\,g_i^2\,\delta(\nu-\nu_i),
\end{equation}
with $\delta(u)$ standing for the Dirac delta-functions. The corresponding $\Phi'(y)$ consists of steps:
$\Phi'(y)=\sum_{i=1}^Kg_i^2\theta(y-\nu_i)$. A simple
consideration shows that our earlier analysis for the values of
$d_{\max}$ and the critical temperature $T_{c}$ still
hold for such a case, so $d_{\max}=R^2/2$, and
$T_{c}=[\Phi'(1)]^{1/2}=\sqrt{g_1^2+g_2^2+\ldots g_K^2}$.
The equation (\ref{minmax}) used to determine $d_{\min}=R^{2y_{\min}}-a^2$ now takes the following form:
\begin{equation}\label{disc1}
T^2=\sum_{i=1}^Kg_{i}^2\frac{1-a^2e^{-y_{\min}/z}}{1+e^{(\nu_i-y_{\min})/z}},
\quad z=\frac{1}{2\ln{R}}.
\end{equation}
A little thought shows that the solution should always be in the
form $y_{\min}=\nu_p+c_pz$ for small $z$, where the index $p$ runs
successively through the values $1,...,K$ when decreasing
temperature from $T_{c}$ towards $T=0$. Introducing a decreasing
sequence of characteristic temperatures
$T_p=\sqrt{\sum_{i=p}^{K}g_i^2}$, we find in the interval
$T_{p+1}<T<T_p$ the value $
y_{\min}=\nu_p+z\ln{(T^2-T_{p+1}^2)/(T_{p}^2-T^2)}$. Thus, the
value of $y_{\min}$ jumps (and thus the size of the smallest
frozen blobs $d_{\min}$)  when crossing each of the temperatures
$T_p$, with the highest one being $T_1=T_c$. It is also clear that
for $\nu_{p} \leq \nu <\nu_{p-1}$ one finds: $x(\nu)=~{T}/{T_p}$
when $T \leq T_p$. Since $T_p$ and $\nu_p$ decrease as $p$
increases, $x(\nu)$ for a given temperature $T < T_c$ is step-wise
constant with jumps at each $\nu_p$; the smaller $\nu$ (i.e. the
smaller the size of the blobs), the larger $x(\nu)$, meaning that
the localisation effect is weaker and finally disappears when
$x(\nu)\to 1$.

The expressions for $(y_{\min},y_{\max})$ suffice to calculate the
free energy expression in the thermodynamic limit. In the
temperature range $T_{p+1}<T<T_p$ we have
\begin{equation}\label{freeendisc}
 -{\cal F}=T\nu_p+2\sum_{i=1}^p(\nu_{i-1}-\nu_i)\,T_i+
\frac{1}{T}\sum_{i=p+1}^K(\nu_{i-1}-\nu_i)\,T^2_i,
\end{equation}
whereas for $T>T_{1}=T_{c}$ the RS expression is
\begin{equation}\label{freeendisc1}
-{\cal F}=T+\frac{1}{T}\sum_{i=1}^K(\nu_{i-1}-\nu_i)\,T^2_i\,.
\end{equation}
Interestingly, these expressions reproduce exactly, {\it mutatis
mutandis} the free-energy of Derrida's GREM \cite{GREM,BoK}, with
a particularly clear interpretation in terms of particle
localization inside smaller and smaller blobs as the temperature
is reduced.

Remembering the mentioned perfect match between the results of
\cite{CLD} and \cite{FS} in the limiting case Eq.(\ref{2c})
it is very tempting to conjecture that the GREM behaviour revealed by
us in the infinite-dimensional setting should also hold in {\it
all} spatial dimensions, down to $N=1$, albeit with the largest
exponent $\nu_{0}<1$. Indeed, essentially the same
mechanisms are at play in both situations.
We hope that the
corresponding RG and travelling wave
formalism of \cite{CLD} can be generalized
to support this conclusion. For finite values of $K$, where
lengthscales are well separated, this looks indeed quite feasible.

If this conjecture is  true, we would then have indeed explicitly
constructed a Parisi landscape in finite dimensions fully in terms
of {\it stationary} Gaussian processes. How do we reconcile this
with the ultrametric properties of the Parisi construction?
Consider the following distance $D_R$ defined for any two points
${\bf r},{\bf r'}$ inside a sphere of the radius $R$ in the
Euclidean space of any dimension:
\begin{equation}\label{distance}
D_R({\bf r},{\bf r'})=\frac{\ln{\left[|{\bf r}-{\bf r'}|^2+a^2\right]}}{2\ln{R}}, \quad 0< |{\bf r}|,|{\bf r'}|\le R
\end{equation}
Parameterizing $|{\bf r}| \equiv R^{\alpha({\bf r})}, \, 0\le
\alpha \le 1$, we see that in fact $\lim_{R\to \infty}D_R({\bf
r},{\bf r'})=\max\{\alpha({\bf r}),\alpha({\bf r'})\}\,. $ The
latter function used as a distance converts the Euclidean sphere
into a so-called ultrametric space: every triangle will have at
least two sides equal. We thus conclude that in our model the
covariance of the random potential depends only on the ultrametric
distance inside our growing sphere, not unlike the original
construction of GREM \cite{GREM,BoK} or directed polymers on a tree
with disordered potential \cite{DS}, cf. discussions in
\cite{2d,CLD} in the single scale case.

Several aspects of the model deserve in our opinion further
investigations, some of them to be discussed elsewhere \cite{prep}.
In particular, the rich behaviour found in the thermodynamics of a
single particle should also have interesting dynamical
counterparts, cf. \cite{BGLD,CuLD}. One also can study
multifractality exponents reflecting the spatial organization of
the Gibbs-Boltzmann weights and the associated singularity
spectrum. Finally, let us mention that in $N=1$ the {\it
mono}scale logarithmic landscape model has in fact deep
connections with the {\it multi}fractal Random Walk construction
suggested in \cite{BMD}. The present model suggests a natural
generalisation to a multiscale logarithmic processes \cite{prep}.

\small{This research was supported by Bessel award from Humboldt
foundation, and by grant EP/C515056/1 from EPSRC (UK). This
project was started during the workshop on Random Matrix Theory
held in Jagellonian University, Cracow, May 2007. We thank the
organisers for this opportunity.}

\end{document}